\documentclass[prb,twocolumn]{revtex4}%
\usepackage{amsfonts}
\usepackage{amsmath}
\usepackage{amssymb}
\usepackage{graphicx}%
\begin{document}
\title[Epicycle method]{An epicycle method for elasticity limit calculations}
\author{Axel van de Walle}
\affiliation{Brown University}
\author{Sara Kadkhodaei}
\affiliation{Brown University}
\author{Ruoshi Sun}
\affiliation{Brown University}
\author{Qi-Jun Hong}
\affiliation{Brown University}

\begin{abstract}
The task of finding the smallest energy needed to bring a solid to its onset
of mechanical instability arises in many problems in materials science, from
the determination of the elasticity limit to the consistent assignment of free
energies to mechanically unstable phases. However, unless the space of
possible deformations is low-dimensional and a priori known, this problem is
numerically difficult, as it involves minimizing a function under a constraint
on its Hessian, which is computionally prohibitive to obtain in low symmetry
systems, especially if electronic structure calculations are used. We propose
a method that is inspired by the well-known dimer method for saddle point
searches but that adds the necessary ingredients to solve for the lowest onset
of mechanical instability. The method consists of two nested optimization
problems. The inner one involves a dimer-like construction to find the
direction of smallest curvature as well as the gradient of this curvature
function. The outer optimization then minimizes energy using the result of the
inner optimization problem to constrain the search to the hypersurface
enclosing all points of zero minimum curvature. Example applications to both
model systems and electronic structure calculations are given.

\end{abstract}
\maketitle

\section{Introduction}

The task of identifying a solid's onset of mechanical instability
\cite{ozolins:unstabrmp,luo:idstrmonb} arises in many problems in materials
science and condensed matter physics, from the determination of the failure
mechanisms \cite{marianetti:grafail} to the consistent assignment of free
energies to mechanically unstable phases \cite{avdw:funstab}.

A complex feature of this problem is that the instability can occur along any
phonon mode and not only along the direction of the applied stress or force.
However, the task of computing the Hessian of the energy surface (i.e.
performing a lattice dynamics calculation) at each level of applied strain
and/or displacements can be computationally demanding. This is especially the
case when electronic structure calculations are used, when the solid
considered has a large unit cell or when a disordered alloy is considered. We
propose a method to determine the point of mechanical instability that is
inspired by the well-known dimer method \cite{jonsson:dimer,voter:hyperdyn}%
\ for saddle point searches but that differs in two respect. First, we propose
a slight modification of the dimer method, which we call the epicycle method,
that provides roughly a factor two improvement in computational efficiency for
the problem of determining the softest phonon mode.\ Second, we embed this
epicycle into an outer-level optimization algorithm that searches for the
lowest energy point that lies at the onset of mechanical instability.

Examples of applications to both model systems and electronic structure
calculations are given. We consider the interesting case of the failure
mechanism of graphene under tension \cite{marianetti:grafail}. We also devote
special attention to the calculation of (free) energies of mechanically
unstable phases \cite{avdw:funstab} and exploit the accuracy of the proposed
method to analyze in detail how the calculated quantities vary smoothly with
composition even through the onset of mechanical instability. We also
demonstrate that different alloy systems which share a common element yield
mutually consistent free energies for that element.

\section{Method}

\subsection{Numerical Method}

\subsubsection{Outline and notation}

In a system of $N$ atoms, let $x$ denote the $3N$ vector of all atomic
positions (and unit cell parameter, if the system is periodic), let $V(x)$
denote the potential energy of the system in that state and let $\kappa(x)$ be
the minimum curvature at $x$, that is, the minimum eigenvalue of the Hessian
(the matrix of second derivatives). Hence, $\kappa(x)>0$ and $\kappa(x)\leq0$
correspond to mechanically stable and unstable regions, respectively. The goal
is to numerically minimize the potential energy $V\left(  x\right)  $ with
respect to $x$, subject to the constraint $\kappa\left(  x\right)  =0$. The
key idea is that the constraint $\kappa(x)=0$ can be maintained by
constraining the system to move perpendicular to direction along which
$\kappa(x)$ varies the fastest. This follows from the fact that the normal to
the hypersurface $\kappa(x)=0$ at $x$ is simply given by the gradient of
$\kappa(x)$. However, it is desirable to avoid the need to compute the
derivative of $\kappa\left(  x\right)  $, which is a\ third derivative of
$V(x)$. In fact, we avoid the need to compute second derivatives as well, via
a modification of the dimer method \cite{jonsson:dimer}. In what follows, we
let $|v|$ denote the Euclidian norm of a vector $v$ and subscripts denote
partial derivatives.

\subsubsection{Inner optimization problem}

\begin{figure}[htb]
\centerline{\includegraphics{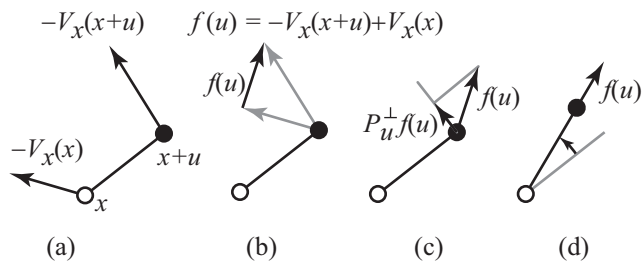}}
\caption{
Epicycle algorithm for the determination of the direction of the most unstable phonon
mode (inner optimization problem). The given structure is specified via the (fixed) vector $x$\ while the direction
of the mode will be determined by the vector $u$ to be optimized.
(a) One first calculates\ the forces $-V_{x}\left( x\right) $, $-V_{x}%
\left( x+u\right) $
at points $x$ and $x+u$, respectively (the force $-V_{x}\left( x\right
) \equiv-g_0 $ needs to be
calculated only once). Next, the difference
$f\left( u\right) \equiv-V_{x}\left( x+u\right) +V_{x}\left( x\right
) $ is calculated (b)
and projected (c) orthogonal to $u$. This orthogonal force $P_{u}^{\perp
}f\left( u\right) $
drives the optimization of the direction $u$, which continues
until (d) the force $f\left( u\right) $ is parallel to $u$.
}
\label{figalgoepi}
\end{figure}

The goal of the inner optimization problem is to find $\kappa(x)$, the minimum
curvature of $V(x)$ at $x$ and can be implemented as follows (refer to Figure
\ref{figalgoepi}). First, compute the gradient at $x$ and set $g_{0}=V_{x}%
(x)$.\ Next, determine the direction $u$ of minimum curvature by minimizing
\[
V(x+u)-g_{0}\cdot u
\]
over all $u$ such that $|u|=\epsilon$, where $\epsilon$ is a user-specified
finite difference step. This procedure works by eliminating the linear term of
Taylor expansion of $V(x)$ through the term $-g_{0}\cdot u$, so that the
remaining quadratic form (up to a $\epsilon^{3}$ error) can be minimized over
a hypersphere to find the direction of minimum curvature. For efficiency
reasons, this minimization is implemented using the gradient of the objective
function, which is equal to $V_{x}(x+u)-g_{0}$. This gradient, projected
orthogonally to $u$, can be used to drive a conjugate gradient optimization
algorithm \cite{fletcher:cg}.

While this step could have been done in the same way as in the original dimer
method \cite{jonsson:dimer,voter:hyperdyn}, we chose here to place the two
images at $x$ and $x+u$ instead of at $x+u$ and $x-u$. The advantage is that
the point $x$ does not move as $u$ is updated, so we only need to recompute
one image at each optimization step, thus essentially halving the
computational requirements. The disadvantage is that a non-central difference
provides a lower order of accuracy. The latter effect can be mitigated at
slight additional cost,\ by using a central difference only at the last step
(or the last few steps). In this work,\ we use a central difference to compute
the minimum curvature after the direction\ $u$ has been optimized with an non-central\ difference.%

\subsubsection{Outer optimization problem}

\begin{figure}[p]
\centerline{\includegraphics{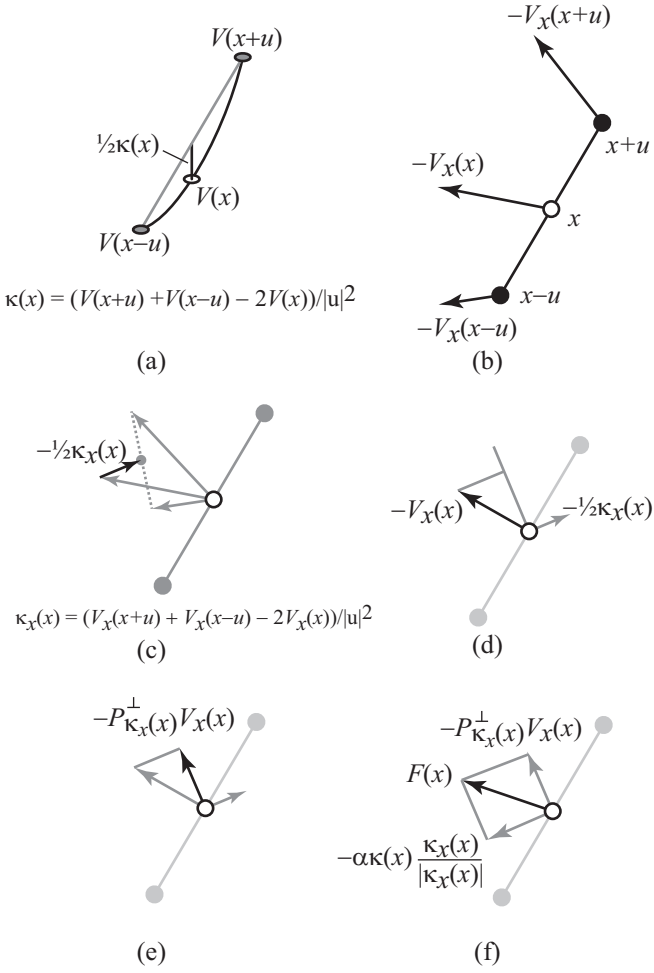}}
\caption{
Computation steps of the outer optimization algorithm.
Grey arrows represent results from the previous steps.
(a) The curvature of the potential $\kappa(x)$ at $x$
can be estimated via the finite difference
$\kappa(x)=\left( V(x+u(x))+V(x-u(x))-2V(x)\right) /\left\vert u\right
\vert^{2}$.
The algorithm also exploits\ the knowledge of the forces (b) acting at points
$x+u$, $x-u$ and $x$ to form the finite difference
$\kappa_{x}(x)=\left( V_{x}(x+u(x))+V_{x}(x-u(x))-2V_{x}(x)\right) /\left\vert
u\right\vert^{2}$,
which yields an estimate of the gradient of the curvature (c).
The force $-V_{x}(x)$ (d) is then projected orthogonally (e) to
$\kappa_{x}(x)$ to yield $-P_{\kappa_{x}\left( x\right) }^{\perp}V_{x}%
\left( x\right) $.
Moving in this direction ensures that the curvature constraint
$\kappa\left( x\right
) =0$ is maintained, to first order, while the energy is being
minimized. (f) To generate an explicit driving force towards points $x$ that satisfy the
constraint $\kappa\left( x\right) =0$, the total
\textquotedblleft force\textquotedblright\ $F\left( x\right
) $ acting on the system is
obtained by adding the contribution from (e) to an attractive force towards the
hyperplane where the constraint is satisfied. The latter is obtained by combining the
curvature information from step (a) and the curvature gradient direction from step (c).
The parameter $\alpha
$ controls the relative strengths of the constraint stringency
force and of the energy minimizing force.
}
\label{figalgoout}
\end{figure}

This optimization problem seeks to minimize energy $V(x)$ under the constraint
of $\kappa(x)=0$ and can be performed as follows (refer to Figure
\ref{figalgoout}). This task will require the knowledge of the gradient of
$\kappa(x)$, denoted $\kappa_{x}(x)$.\ To find a convenient expression for it,
note that we can express $\kappa(x)$ via finite differences as $\kappa
(x)=(V(x+u(x))+V(x-u(x))-2V(x))\epsilon^{-2}$, where $u(x)$ is the solution to
the inner optimization problem at $x$. Since the function $\kappa(x)$ has
already been optimized with respect to $u$, calculating this derivative does
not need to account for changes in $u(x)$ (a well-known result from
optimization theory that is used, for instance, in first-order perturbation
theory).\footnote{To see this directly, letting $g\left(  x\right)  $ denote
the value of $u$ that minimizes $f\left(  x,u\right)  $ for a given $x$, we
have $\partial f\left(  x,u\right)  /\partial u=0$ at $u=g\left(  x\right)  $
(assuming smoothness and no boundary solution) and thus%
\[
\frac{df\left(  x,g\left(  x\right)  \right)  }{dx}=\frac{\partial f\left(
x,g\left(  x\right)  \right)  }{\partial x}+\frac{\partial f\left(  x,g\left(
x\right)  \right)  }{\partial u}\frac{\partial g\left(  x\right)  }{\partial
x}=\frac{\partial f\left(  x,g\left(  x\right)  \right)  }{\partial x}.
\]
} The gradient of $\kappa(x)$ thus admits, to first order,\ a very simple
expression that only involves gradients of $V(x)$:
\[
\kappa_{x}(x)=\left(  V_{x}(x+u(x))+V_{x}(x-u(x))-2V_{x}(x)\right)
\epsilon^{-2}.
\]

If one happens to start the optimization from a point such that $\kappa(x)=0$,
then it is sufficient to move $x$ in the direction opposite to the gradient
$V_{x}(x)$, projected orthogonal to $\kappa_{x}(x)$. This ensures that the
constraint $\kappa(x)=0$ remains satisfied (to first order) as the energy is
being minimized. During a numerical optimization, however, the update steps
are not infinitesimal, hence $\kappa(x)$ will gradually deviate from zero as
the optimization progresses. To avoid this, we add a force, parallel to
$\kappa_{x}(x)$, proportional in magnitude to $\kappa(x)$ and in a direction
such that it brings the system back towards the hypersurface where
$\kappa(x)=0$. This additional force also has the desirable side-effect that
the constraint does not need to be already satisfied at the starting point of
the optimization, since the method generates an attractive force towards the
$\kappa(x)=0$ hypersurface.

The force acting on the system is then given by the sum of these two
contributions:
\begin{equation}
F(x)=-P_{\kappa_{x}(x)}^{\perp}V_{x}(x)-\alpha\kappa(x)\frac{\kappa_{x}%
(x)}{|\kappa_{x}(x)|} \label{eqfout}%
\end{equation}
where $P_{v}^{\perp}=I-P_{v}^{\parallel}$, in which $P_{v}^{\parallel}%
=v(v^{T}v)^{-1}v^{T}$ is a matrix that projects onto the vector $v$, while
$\alpha$ is a constant controlling the strength of the attraction to the
constraint hyperplane. The force $F(x)$ can be fed into a standard
gradient-driven optimization routine. Note that one can also arrive at
Equation (\ref{eqfout}) via a standard Lagrange multiplier argument.

It is instructive to write the standard dimer method in a similar notation to
emphasize the differences:%
\begin{equation}
F\left(  x\right)  =-P_{u(x)}^{\perp}V_{x}(x)+P_{u(x)}^{\parallel}V_{x}(x)
\label{eqfdim}%
\end{equation}
where $u\left(  x\right)  $ is the direction of the dimer. Note that the
gradient $V_{x}(x)$ is projected onto $u\left(  x\right)  $ in the dimer
method instead of $\kappa_{x}\left(  x\right)  $ in our method. Also, the
force along $\kappa_{x}(x)$ is determined by a simple projection of
$V_{x}\left(  x\right)  $ onto $u$ in the dimer method rather than being
jointly determined by the curvature $\kappa\left(  x\right)  $ and its
gradient $\kappa_{x}\left(  x\right)  $.%

\begin{figure}[htb]
\centerline{\includegraphics{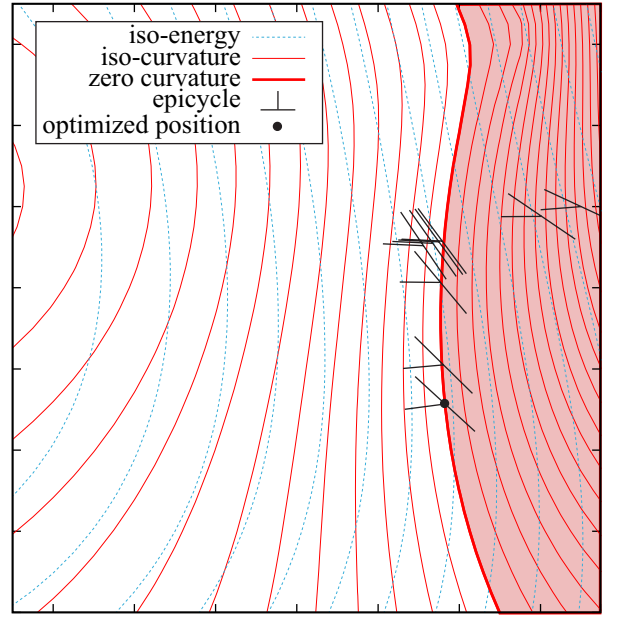}}
\caption{\label{figlc}%
(Color online) Convergence of the proposed ``inflection detection'' method for
the determination of inflection point. The figure shows contours of constant energy
(which should be minimized) and contours of constant curvature (which is a constraint)
for a test case involving a highly nonlinear analytic function.
The dimer is shown as a ``T'' whose top line indicates the direction of smallest curvature
and whose vertical line indicates the direction of largest curvature change.
Conjugate gradient optimization steps are shown.
The optimized position is seen to lie on a line of zero curvature at the point of
minimum energy within the shaded region (in which the minimum curvature is negative).}
\end{figure}%

Figure \ref{figlc} shows a calculated inflection point in a simple analytic
example (chosen to be two-dimensional, to enable a graphical representation)
with the potential $V\left(  x,y\right)  =\left(  x^{2}+y^{2}\right)
/2-\left(  x^{2}+y^{2}\right)  ^{3/2}/6+\sqrt{y}+x/8+3\exp\left(  -\left(
x-23/10\right)  ^{2}-\left(  9/16\right)  \left(  y-0.5\right)  ^{2}\right)
+x^{2}+y/5.$

\subsubsection{Implementation details}

\label{secdetails}A few aspects of the implementation deserve some attention.
First, in the common situation of a periodic system, the state vector $x$ must
also contain the degrees of freedom corresponding to the unit cell shape. To
ensure that, regardless of the number of atoms in the unit cell, the entries
in the variable $x$ have comparable magnitudes and the corresponding forces
have comparable magnitude as well, we use the following scaling. The strain
$\varepsilon_{ij}$ applied on the unit cell is stored in $x$ as a scaled
strain $\tilde{\varepsilon}$, defined as
\begin{equation}
\tilde{\varepsilon}_{ij}=\gamma^{-1}n\Omega^{1/3}\varepsilon_{ij},
\label{eqscstrain}%
\end{equation}
where $\Omega$ is the average volume per atom, $n$ is the number of atoms in
the unit cell and $\gamma$ is a dimensionless user-specified parameter. The
corresponding scaled stress $\tilde{\sigma}$ is given by%
\begin{equation}
\tilde{\sigma}_{ij}=\gamma\Omega^{2/3}\sigma_{ij}, \label{eqscstress}%
\end{equation}
where $\sigma_{ij}$ is the actual stress.\ One can readily verify that the
product of the two scaled conjugate variables is, as it should,\ $\gamma
^{-1}n\Omega^{1/3}\varepsilon_{ij}\gamma\Omega^{2/3}\sigma_{ij}=V\varepsilon
_{ij}\sigma_{ij}$, where $V=n\Omega$ is the cell volume. This convention
offers the advantage that the scaled stress $\Omega^{2/3}\sigma_{ij}$ has
units of force and its magnitude is independent of the cell size (since
$\Omega$ is the volume per atom and not per cell). Also, the scaled strain
$n\Omega^{1/3}\varepsilon_{ij}$ has units of length and ensures that, as the
unit cell size $n$ increases, a given amount of scaled strain corresponds to a
smaller actual strain. As a result, the change in energy corresponding to a
given level of scaled strain does not grow with unit cell size.

A related issue is that, ideally, the atomic coordinates and forces should
also be independent of cell size. Consequently, it is inconvenient to use
fractional coordinates in the vector $x$ because a given level of scaled force
could be associated with very different real magnitude of the forces if the
unit cell is noncubic. To avoid this, we directly store the atoms' cartesian
coordinates in the vector $x$. We use their coordinates before the whole
system is uniformly strained by the strain $\varepsilon$ (as this choice
avoids complex coupling terms in the stress tensor).\ These scalings offer the
advantages that the tolerance criterion for convergence can be specified in
easy-to-interpret units and that it does not need to be adjusted for cell size.

It is important to realize that the method is only able to identify an
unstable mode that can be represented with the supercell considered. It is
possible that, once the onset of instability for a given supercell has been
found, a full lattice dynamics analysis would reveal an unstable mode
involving correlated motion over a cell bigger than the one considered. In
such case, one can simply create a supercell that has the right size and shape
in order to accommodate the unstable phonon mode (if there are multiple, it is
advisable to consider the most unstable mode) and re-run our method on that
supercell. Even though a lattice dynamics calculation is still necessary, we
still avoid the need to perform a large number of such calculations, once for
each trial geometry, thus considerably reducing the computational burden.

Equation (\ref{eqfout}) involves a user-specified stiffness parameter $\alpha$
(which can also be viewed as Lagrange multiplier). The method is theoretically
valid for any value of $\alpha$, but different values merely weigh differently
the stringency of the constraint versus the accuracy of the minimum energy. In
practice, we set it to a reasonable value such that the two terms of Equation
(\ref{eqfout}) have the same magnitude for the initial trial value of $x$. We
then leave its value unchanged for the remaining iterations. This choice
typically results in a fairly well conditioned optimization problem (with all
terms in Equation (\ref{eqfout}) having similar orders of magnitudes).
Although one might think that the value of $\alpha$ could be set automatically
if one formally\ solved the constrained optimization problem via the Lagrange
multiplier method, this is not\ the case: Multiplying the constraint
$\kappa\left(  x\right)  =0$ by an arbitrary constant changes the conditioning
of the Lagrangian optimization problem in the same way as changing the factor
$\alpha$ does in our approach.

A few other important points that apply equally to the original dimer method
should be kept in mind. The inner optimization problem is very well behaved
because it corresponds to minimizing a quadratic form on a hypersphere. As a
result, conjugate gradient methods perform very well and are quite robust.
However, the outer optimization problem can be substantially more nonlinear.
In addition, it is important to realize that the force obtained by Equation
(\ref{eqfout}) is not, in general, guaranteed to be the exact gradient of some
function. It does behave as a proper gradient in the limit of approaching the
solution, however.\ These observations suggest that optimization methods that
require function values (in addition to gradients) are not well suited to
drive the outer optimization problem. Similarly, methods that rely on a
quadratic form assumption should only be used for refinement, once a
reasonably good solution has already been obtained. In our experience,
performing a few steps of descent along the gradient is a good way to improve
the initial guess of the solution before iterating a conjugate gradient
algorithm to convergence.

An efficient implementation of the method should exploit the fact that the
epicycle typically does not need to turn very much between two steps of the
outer optimization algorithm (as is visible in Figure \ref{figlc}). That is,
the direction of the most unstable mode varies smoothly with the structure's
geometry. As a result, the first invocation of the inner optimization problem
will typically take about as long as a standard structural relaxation in order
to find the most unstable mode but subsequent invocation will typically only
demand the equivalent of a few static calculations.

To reach a given energy accuracy, the number of steps in the outer
optimization routine tends to be slightly larger than the number of steps
needed in a standard structural relaxation. This is due to the fact that the
energy is quadratic in the atomic displacements near a minimum while the
energy is linear in those displacements near an inflection point. As a result,
the energy is less sensitive to a precise structural optimization near a
minimum than near an inflection point and fewer optimization steps are thus
needed in the former problem than in the latter. The net effect is that the
computational cost of the proposed method is just a few times larger than the
cost of a standard energy minimization.

When interfacing the method with a given total-energy ab initio\ code, a few
more practical issues need some attention.\ First, it is crucial that all
electronic structure calculations be carried out with constant basis set and
the same $k$-points. Otherwise some of the numerical derivatives fail to
behave smoothly, which may confuse the numerical optimization routines.
Second, considerable efficiency improvements can be achieved if the ab initio
code can exploit charge density prediction and re-use previously converged wavefunctions.

The above algorithm has been implemented as a C++ code and is now included in
the Alloy Theoretic Automated Toolkit (ATAT)
\cite{avdw:maps,avdw:atat,avdw:atat2} (the key commands are \texttt{infdet}
and \texttt{robustrelax\_vasp}). The algorithm can thus easily be interfaced
with any of the ab initio codes supported by ATAT. In the present work, we
used the VASP ab initio code \cite{kresse:vasp1,kresse:vasp2,kresse:paw} and
the interface to that code is more developed. For instance, helper routines
automatically prepare some of the input files, copy files to re-use
pre-converged results, ensure that a constant basis set is used (the present
implementation of VASP's constant basis set restarts require the user to
specify the plane wave basis explicitly --- our interface takes care of
this.), etc.

\section{Examples}

All structural optimizations reported below are driven\ by forces obtained
from electronic structure calculations performed with VASP
\cite{kresse:vasp1,kresse:vasp2,kresse:paw} in conjunction with the
ATAT\ package \cite{avdw:atat,avdw:atat2,avdw:mcsqs} to model random solid
solutions and the Phonopy software \cite{togo:phonopy} to calculate phonon
spectra.\ More computational details are given in the Appendix.

\subsection{Failure mechanism of graphene under tension}

\label{secgraphene}The behavior of graphene under tension at the limit of its
ideal strength has recently been studied in detail \cite{marianetti:grafail}
with the unexpected finding that the mechanical instability does not develop
along the direction in which the strain is applied but instead involves the
appearance of an unstable optical mode. This unexpected finding provides a
compelling example of the usefulness of our approach, as it is specifically
designed to efficiently explore all possible instabilities without performing
a full lattice dynamics analysis at each trial configuration.%

\begin{figure}[htb]
\centerline{\includegraphics{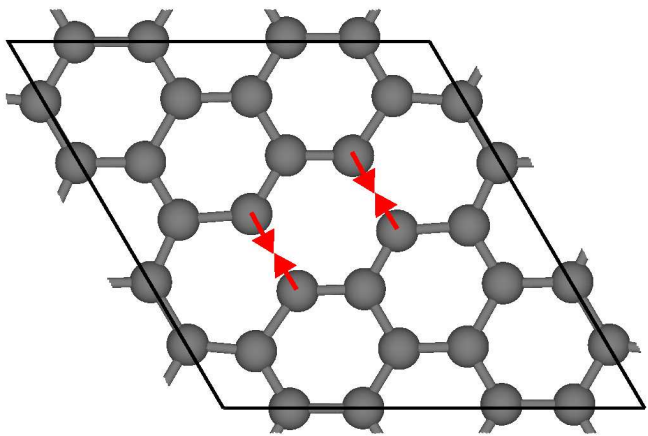}}
\caption
{(Color online) Geometry of graphene at the limit of mechanical stability.
Unit cell shown by a thick outline.
Arrows indicate the most significant displacements associated with the unstable mode.
}
\label{figgraphon}
\end{figure}%
%

\begin{figure}[htb]
\centerline{\includegraphics{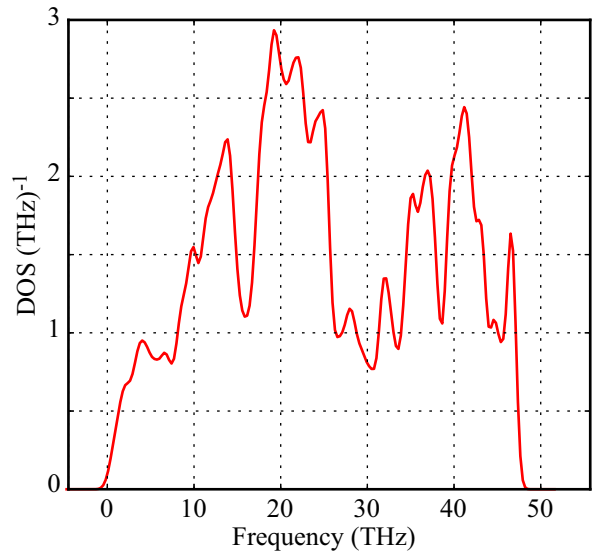}}
\caption
{(Color online) Phonon Density of States (DOS) of graphene at the limit of its mechanical stability.
}
\label{figgraphdos}
\end{figure}%

Our method provides a different perspective on this phenomena rather than
merely corroborating the earlier finding. Our approach accounts for the fact
that above absolute zero the system would visit many states in the
neighborhood of its ideal high-symmetry crystal structure. Consequently, the
point of mechanical instability would not necessarily be uniformly strained
version of graphene's ideal crystal structure. We find that, if thermal noise
is accounted for, the system quickly breaks its symmetry and an unstable mode
develops at a low strain (principal strains: 3.1\% and 4.6\%) about a
low-symmetry structure rather than at a higher strain about a high-symmetry
structure. The method delivers the point of lowest energy that lies on the
edge of the domain of mechanical stability, thus discovering the
\textquotedblleft weakest link\textquotedblright\ of the material's stability,
instead of\ finding the location where a phonon instability develops along a
given pre-specified path. Note that one can recover the earlier results
\cite{marianetti:grafail} by simply constraining the outer optimization
problem to only explore isotropically strained version of the ideal graphene structure.

To identify the onset of instability, we first ran the method on a small
supercell of 6 atoms, selected to match the supercell of the known unstable
mode identified earlier \cite{marianetti:grafail}. Our method found an onset
of instability for a mode that breaks exactly one bond per supercell of 6
atoms. A phonon analysis of this structure revealed an unstable phonon branch
with a minimum at the Brillouin zone boundary. We thus created a supercell of
24 atoms that could represent this most unstable mode. Our method then found
the structure depicted in Figure \ref{figgraphon} and a lattice dynamics
analysis no longer indicated any unstable modes (see Figure \ref{figgraphdos}%
). The \textquotedblleft weakest link\textquotedblright\ mode identified takes
the form of the simultaneous breaking of two nearby bonds, repeated in a
periodic pattern that appears to minimize the distortion of other bonds. This
geometry could not have been anticipated from simple geometric or chemical
arguments, which illustrates the usefulness of the method.

\subsection{Free energy of mechanically unstable phases}

\begin{figure}[htb]
\centerline{\includegraphics{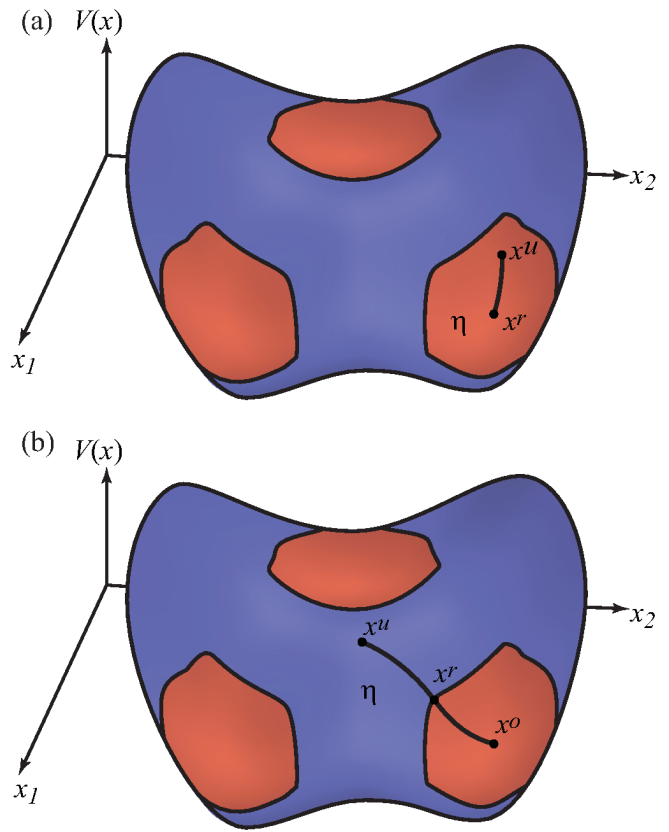}}
\caption{(Color online)
Inflection-detection method. The potential energy hypersurface $V(x)$ (as a function
of the state $x$ of the system) defines a natural partitioning of phase space into
neighborhoods $\eta$, based on the sign of $\kappa\left( x\right) $, the local
minimum curvature of $V(x)$ (blue: negative, red: positive). Each neighborhood
(stable or not) can be assigned a well-defined energy by finding the minimum
energy within that neighborhood. (a) In the case of a mechanically stable structure,
the initial unrelaxed structure $x^{u}%
$ simply relaxes to a local minimum $x^{r}$.
(b) For a mechanically unstable structure, an unconstrained minimization would
yield the over-relaxed point $x^{o}%
$ which is actually the energy of another structure.
The inflection detection method instead finds $x^{r}%
$, the minimum energy within
$\eta$, which is located at an inflection point where the minimum curvature
$\kappa\left( x\right) $ changes sign.
}
\label{figpotsurf}
\end{figure}

It has recently been shown \cite{avdw:funstab} that the point of lowest energy
at the onset of mechanical instability (as identified with the proposed
method) provides a logically consistent definition of the energy of a
mechanically unstable phase. For completeness and clarity, we summarize the
main features of this approach\ below (refer to Figure \ref{figpotsurf}). The
set of coordinates $x$ such that\ $\kappa(x)>0$ and $\kappa(x)\leq0$
correspond to mechanically stable and unstable regions, respectively. Given an
ideal structure $x^{u}$ in which atoms are not allowed to relax away from
their ideal positions, we define its neighborhood $\eta$ as the largest
connected set containing $x^{u}$ over which the minimum curvature $\kappa(x)$
does not change sign. Now, we define the energy $E$ associated with $x^{u}$ as
the minimum of the potential $V\left(  x\right)  $ over all $x$ in the
neighborhood $\eta$. When $x^{u}$ is in a mechanically stable region, $E$ is
just the potential energy $V\left(  x^{r}\right)  $ at the lowest local
interior minimum $x^{r}$ in $\eta$, which agrees with the usual notion of
energy of a relaxed structure. When $x^{u}$ is in a mechanically unstable
region, $E=V\left(  x^{r}\right)  $ as well, but now $x^{r}$ must be at the
boundary of $\eta$, i.e., a point where $\kappa(x)$ is zero. Specifically,
$x^{r}$ is the point of minimum energy subject to the constraint that
$\kappa\left(  x^{r}\right)  =0$. This definition offers three desirable
properties: (i) it is based on a simple geometrical\ notion of curvature (ii)
it can be shown that, at the onset of mechanical instability, a local minima
always merges with a point of zero minimum curvature, so energy is a
continuous function even across an instability and (iii) the relaxed structure
$x^{r}$ is such that it is only unstable along at most a small finite number
of modes (which is negligible in the thermodynamic limit of an infinite number
of modes), so a standard lattice dynamics calculation can be used to calculate
the free energy. This method is called \textquotedblleft inflection
detection\textquotedblright\ because $x^{r}$ lies at an inflection point for a
mechanically unstable structure.%

%

\begin{figure}[htb]
\centerline{\includegraphics{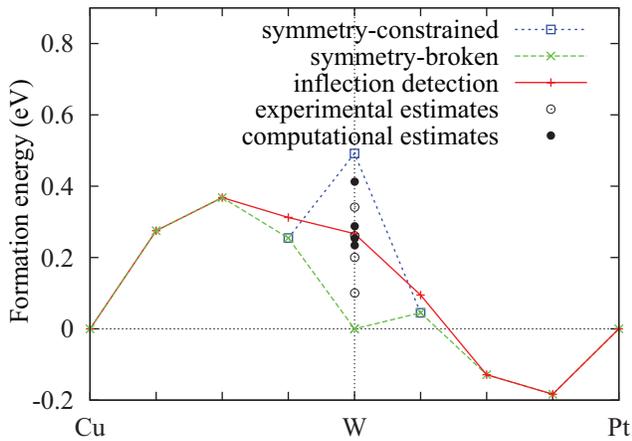}}
\caption{(Color online)
Composition-dependence of the formation energy in the Cu-W and Pt-W fcc alloy systems.
Symmetry-constrained results employ structural relaxations that preserve the initial
symmetry of the structure while symmetry-broken ones allow the system to relax
to its unconstrained minimum energy.
Inflection-detection results represent either a local minimum or an inflection point energy,
depending on which one is closest to the initial unrelaxed structure.
The inflection-detection graph is seen to be the only one that generates a smooth
behavior across all compositions that furthermore converges to a common value
for the element W that both alloy systems share. This value compares favorably
with available experimental\cite
{guillermet:ptw,saunders:metastab,sgte:db,kaufman:tc}
and computational\cite{ozolins:w,avdw:funstab} estimates.
}
\label{figcuptw}
\end{figure}%
%

\begin{figure}[htb]
\centerline{\includegraphics{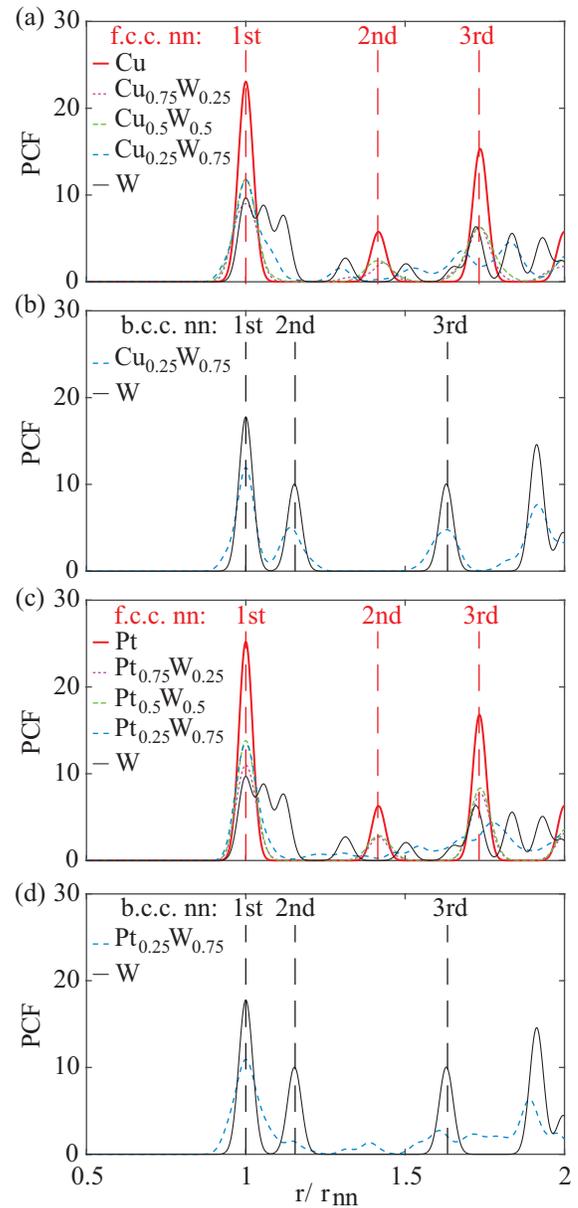}}
\caption{(Color online)
Pair correlartion functions (PCF) for various disordered alloy systems:
(a) fcc Cu$_x$W$_{1-x}$,
(b) bcc Cu$_x$W$_{1-x}$,
(c) fcc Pt$_x$W$_{1-x}$ and
(d) bcc Pt$_x$W$_{1-x}$.
For fcc structures with $x \leq
0.25$, the structure geometries correspond to the inflection point determined with the proposed algorithm.
In all other cases, the structure geometries correspond to local minima.
}
\label{figpcfcuptw}
\end{figure}%
%

\begin{figure}[htb]
\centerline{\includegraphics{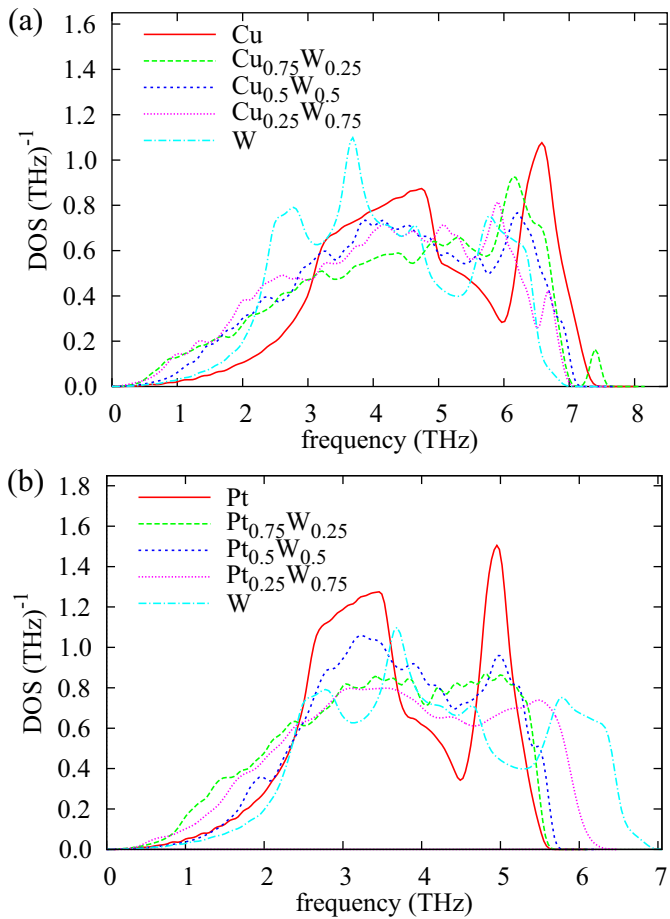}}
\caption{(Color online)
Composition-dependence of the phonon Density of States (DOS)
in the (a) Cu-W and (b) Pt-W fcc alloy systems.
}
\label{figdosCuPtW}
\end{figure}%
%

\begin{figure}[htb]
\centerline{\includegraphics{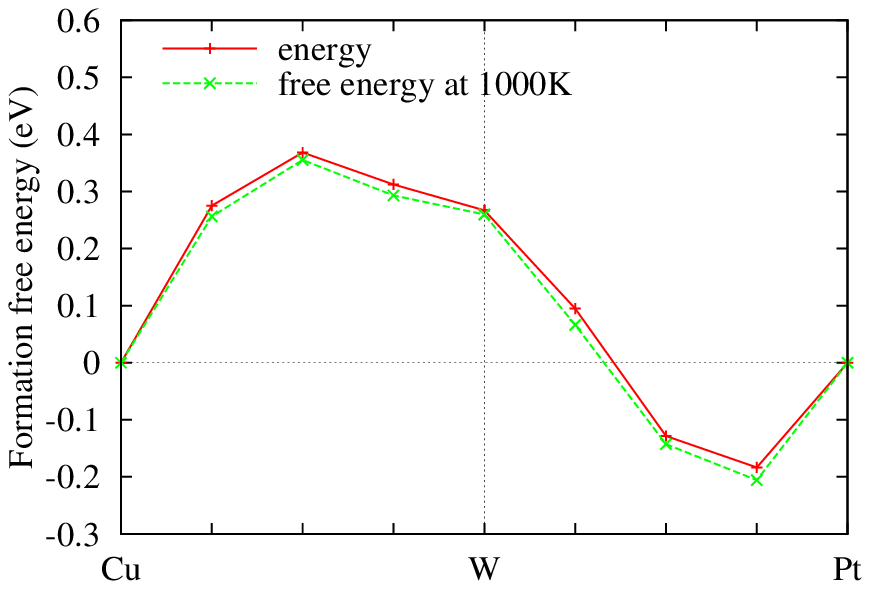}}
\caption{(Color online)
Composition-dependence of the formation energy and free energy
(at 1000 K) in the Cu-W and Pt-W fcc alloy systems.
}
\label{figwphon}
\end{figure}%

The inflection point calculations presented in \cite{avdw:funstab} relied on
the Nudged Elastic Band Method (NEB) \cite{jonsson:nudged},\ which has two
drawbacks in the context of inflection point calculations. First, this
approach demands rather large-scale calculations in which multiple copies of
the whole system are being simultaneously optimized to map out the path along
the elastic band. Second, this method is only sensitive to instabilities that
develop along the path and not perpendicular to it. In contrast, the method
proposed here enables the efficient calculation of the inflection point using
only one image of the whole system and is sensitive to instabilities along any
of the phonon modes (that can be represented within the supercell used).

In this application, we have found that a reasonable starting point for the
algorithm can be obtained from the midpoint between a fully relaxed structure
(allowing cell parameters and ionic positions to vary) and a constrained
relaxation in which only the isotropic changes in the unit cell are allowed
(i.e. a \textquotedblleft volume only\textquotedblright\ relaxation).

As an illustration, we compute formation energies of mechanically unstable
phases fcc solid solutions in the Cu-W and Pt-W systems, which share the W
component, whose stable structure is bcc, while the stable structure of Cu and
Pt is fcc. The disordered alloys in these systems are modeled via Special
Quasirandom Structures (SQS) \cite{zunger:sqs} generated with the
\texttt{mcsqs} code \cite{avdw:mcsqs} of the ATAT package
\cite{avdw:atat,avdw:atat2}.

Figure \ref{figcuptw} shows formation energies (relative to a phase-separated
mixture of fcc Cu, fcc Pt and bcc W of the same overall composition) as a
function of composition computed in various ways. The symmetry-constrained
result is obtained by performing structural relaxations that preserve the
initial symmetry of the structure before the relaxation steps (which is the
default behavior for most ab initio codes). The symmetry-broken results are
equivalent to the symmetry-constrained ones, except for mechanically unstable
high-symmetry structures (here fcc W), where the symmetry is explicitly broken
to allow the system to relax to its unconstrained minimum energy (here bcc W).
The latter scheme avoids mechanical instabilities but has the undesirable
consequence that fcc W is actually assigned the energy of bcc W.\ Finally, the
inflection detection curve is obtained by minimizing energy until one finds
either a local minimum or an inflection point, and reporting the energy of
whichever is first found. (Another way to describe the approach is to state
that one uses fully relaxed energies for all structures that maintain an
fcc-type coordination upon relaxation and inflection point energies for
structures that do not.)

In the Cu-W binary, the fcc structure is thermodynamically unstable at all but
very dilute compositions (as indicated by the positive formation energies).
However, the fcc structure is nevertheless mechanically stable at least up to
50 atomic \% W. The data point at 75 atomic \% W clearly exhibits mechanical
instability (since the relaxed and inflection detection results differ). This
is also visible in Figure \ref{figpcfcuptw}, which shows the pair correlation
functions (PCF) for each structure. For Cu, Cu$_{0.75}$W$_{0.25}$ and
Cu$_{0.5}$W$_{0.5}$, the PCF is clearly indicative of an fcc structure (see
Figure \ref{figpcfcuptw}(a)). For Cu$_{0.25}$W$_{0.75}$ and W, the fully
relaxed structures exhibit a bcc-like PCF, as shown in Figure
\ref{figpcfcuptw}(b). For these structures, we thus use the inflection
detection method and the resulting PCF, shown in Figure \ref{figpcfcuptw}(a),
are almost fcc-like: The first nearest neighbor peak is at the same distance,
although it is a bit broadened, especially for W. The second nearest neighbor
peak splits into two peaks whose centers\ average to the corresponding fcc peak.

In the Pt-W binary, the fcc structure is mechanically stable for a broad range
of compositions, as can be seen from the fact that all three methods agree for
most data points in Figure \ref{figcuptw} and the fact that the corresponding
PCF, shown in Figure \ref{figpcfcuptw}(c) are fcc-like. The points at 75
atomic \% W is relaxing to a bcc-like structure (as indicated by the PCF shown
in Figure \ref{figpcfcuptw}(d), although the second nearest neighbor peaks are
not as well defined as for Cu$_{0.25}$W$_{0.75}$). For this point, the
inflection detection method also yields a fcc-like PCF (see Figure
\ref{figpcfcuptw}(c)) similar to Cu$_{0.25}$W$_{0.75}$.

In Figure \ref{figcuptw}, we show the composition dependence of the energy in
both systems on a common graph to demonstrate that the energies of both alloys
smoothly converge to an identical value for the common element W in the
(virtual) fcc structure. Interestingly, this value of the energy of fcc W is
very consistent with previous estimates
\cite{guillermet:ptw,saunders:metastab,sgte:db,kaufman:tc,ozolins:w}: It falls
roughly in the middle of the cloud of estimates and at a location where there
is an increased density of earlier data points. It is interesting to observe
that, if one only looked at the Pt-W system, one may be led to believe that
using symmetry-constrained energy is the \textquotedblleft
right\textquotedblright\ approach while if one only looked at the Cu-W system,
one would think that using symmetry-broken energies is the \textquotedblleft
right\textquotedblright\ approach. But these two approaches would yield
different values for the W energy. In contrast, the inflection detection
approach, which also yields a smooth behavior for both systems, converges to a
common value for the pure element W for both systems.

The inflection detection method also guarantees that all phonon modes but one
are stable, a property that can be independently verified by lattice dynamics
calculations (see Figure \ref{figdosCuPtW}). This property enables the
calculation of phonon free energy contributions via a standard harmonic
treatment. As shown in Figure \ref{figwphon}, the resulting free energies also
vary smoothly with composition, reflecting the relatively smooth
composition-dependence of the DOS (see Figure \ref{figdosCuPtW}). The free
energies as a function of composition also converge to a common value for pure
W, as seen in Figure \ref{figwphon}. Note that the configurational entropy is
deliberately excluded in this graph because its singular behavior (of the form
$X\ln X$ as $X\rightarrow0$) near pure compositions would mask the smoothness
of the phonon contributions.

\section{Conclusion}

We have devised a formal method to determine the point of minimum energy lying
at the onset of mechanical stability. Exploiting some of the same principles
underlying the well-known dimer method, our approach avoids the need to
compute higher order derivatives of the energy (as only forces are needed).
However, our method differs in nontrivial ways from the dimer method because
it seeks minimum energy inflection points rather than saddle points.

Our method proves useful in investigating the mechanisms for mechanical
failure at the atomic level, as illustrated in the case of graphene.\ Our
example of application to Cu-W and Pt-W disordered alloys also support the
proposal \cite{avdw:funstab} that the determination of the lowest energy
inflection point provides a reliable, well-defined and computationally
convenient way to assign energies to mechanically unstable phases. The
proposed method makes this approach even more attractive, because the
computational cost associated with finding the inflection point is just a few
times larger than that of a standard structural relaxation. An implementation
of the proposed method is now part of the ATAT package
\cite{avdw:atat,avdw:atat2} (as the command \textquotedblleft\texttt{infdet}%
\textquotedblright\ or the script \textquotedblleft\texttt{robustrelax\_vasp}%
\textquotedblright).

\section*{Acknowledgments}

This work is supported by the US Office of Naval Research via grant
N00014-14-1-0055 and by Brown University through the use of the facilities of
its Center for Computation and Visualization. This work uses the Extreme
Science and Engineering Discovery Environment (XSEDE), which is supported by
National Science Foundation grant number ACI-1053575.

%

\appendix*%

\section{Computational Methods Details}

Electronic structure calculations are performed with VASP
\cite{kresse:vasp1,kresse:vasp2,kresse:paw} using the PBE exchange-correlation
functional \cite{pbe:pbe} and Projector-Augmented Wave
(PAW)\ \cite{blochl:paw} pseudopotentials. The precision flag is set to
\textquotedblleft high\textquotedblright\ (which specifies the kinetic energy
cutoff) while the number of $k$-point is set to $4000$ per reciprocal atom
(see \cite{avdw:maps} for a description of this convention) with a Gaussian
smearing of 0.1 eV (for force calculations) and the tetrahedron method with
Bl\"{o}chl corrections for total energies. The energy convergence criterion of
$10^{-5}$ eV is used.

The SQS used to model random solid solution were generated with the mcsqs tool
\cite{avdw:mcsqs} of the ATAT\ package \cite{avdw:atat,avdw:atat2}. The SQS
used are those\ included in the package's distribution and have a unit cell of
32 atoms. These SQS match the pair correlations of the disordered state up to
third nearest neighbor as well as the triplet correlations at least up to the
second nearest neighbor.

The epicycle length for the inflection detection algorithm is set to
$\epsilon=$ $0.2$ \AA ~and the user-specified factor for scaling the strain
and stress in Equations (\ref{eqscstrain}) and (\ref{eqscstress}) is
$\gamma=3$. The parameter $\alpha$ in Equation (\ref{eqfout}) is set
automatically, as described in Section \ref{secdetails} and its value ranged
from $\alpha\approx3$ to $\alpha\approx200$ in our applications, depending of
the system. The method was iterated to yield an energy accuracy better than
$1$ meV/atom and a curvature accuracy better than $0.02$ eV/\AA $^{2}$.

Lattice dynamics calculations are performed with the Phonopy software
\cite{togo:phonopy}. For the lattice dynamics calculations of graphene, 144
force calculations are performed on a 96-atom unit cell (a $2\times2\times1$
supercell of the 24-atom cell mentioned in Section \ref{secgraphene}) with a
$k$-point mesh of $3\times3\times1$ and imposed displacement of 0.01\ \AA .
For each displacement, a displacement in the exact opposite direction is also
considered, to cancel out the effect of any nonzero gradient on the
calculation of the Hessian. A phonon $k$-point mesh of $11\times11\times1$ is
used. For the lattice dynamics calculations of Cu-W and Pt-W alloys, the force
calculations employ symmetrically distinct displacements of 0.015 \AA . For
simple fcc or bcc structures, a $3\times3\times3$ supercells is used while for
SQS the supercell is either the structure's unit cell (for large 32-atom SQS),
or, in some cases (Cu$_{0.25}$W$_{0.75}$ and Pt$_{0.25}$W$_{0.75}$), a
$2\times1\times1$ supercell (consisting of 64 atoms) of the SQS unit cell.\ A
phonon $k$-point mesh of $21\times21\times21$ is used.

\end{document}